\documentclass[twocolumn,showpacs,preprintnumbers,amsmath,amssymb]{revtex4}
\usepackage[dvips]{graphicx}
\newcommand{\thp}{$\Theta^+$ }
\begin{document}
%\draft{}
%
\bibliographystyle{try}

\topmargin -0.5cm

\newcounter{univ_counter}
\setcounter{univ_counter} {0}

\addtocounter{univ_counter} {1} 
\edef\JLAB{$^{\arabic{univ_counter}}$ } 

\addtocounter{univ_counter} {1} 
\edef\OHIOU{$^{\arabic{univ_counter}}$ } 

\addtocounter{univ_counter} {1} 
\edef\ASU{$^{\arabic{univ_counter}}$ } 

\addtocounter{univ_counter} {1} 
\edef\CMU{$^{\arabic{univ_counter}}$ } 

\addtocounter{univ_counter} {1} 
\edef\SCAROLINA{$^{\arabic{univ_counter}}$ } 

\addtocounter{univ_counter} {1} 
\edef\SACLAY{$^{\arabic{univ_counter}}$ } 

\addtocounter{univ_counter} {1} 
\edef\UCLA{$^{\arabic{univ_counter}}$ } 

\addtocounter{univ_counter} {1} 
\edef\CUA{$^{\arabic{univ_counter}}$ } 

\addtocounter{univ_counter} {1} 
\edef\CNU{$^{\arabic{univ_counter}}$ } 

\addtocounter{univ_counter} {1} 
\edef\UCONN{$^{\arabic{univ_counter}}$ } 

\addtocounter{univ_counter} {1} 
\edef\DUKE{$^{\arabic{univ_counter}}$ } 

\addtocounter{univ_counter} {1} 
\edef\EDINBURGH{$^{\arabic{univ_counter}}$ } 

\addtocounter{univ_counter} {1} 
\edef\FIU{$^{\arabic{univ_counter}}$ } 

\addtocounter{univ_counter} {1} 
\edef\FSU{$^{\arabic{univ_counter}}$ } 

\addtocounter{univ_counter} {1} 
\edef\GWU{$^{\arabic{univ_counter}}$ } 

\addtocounter{univ_counter} {1} 
\edef\GLASGOW{$^{\arabic{univ_counter}}$ } 

\addtocounter{univ_counter} {1} 
\edef\INFNFR{$^{\arabic{univ_counter}}$ } 

\addtocounter{univ_counter} {1} 
\edef\INFNGE{$^{\arabic{univ_counter}}$ } 

\addtocounter{univ_counter} {1} 
\edef\ORSAY{$^{\arabic{univ_counter}}$ } 

\addtocounter{univ_counter} {1} 
\edef\ITEP{$^{\arabic{univ_counter}}$ } 

\addtocounter{univ_counter} {1} 
\edef\JMU{$^{\arabic{univ_counter}}$ } 

\addtocounter{univ_counter} {1} 
\edef\KYUNGPOOK{$^{\arabic{univ_counter}}$ } 

\addtocounter{univ_counter} {1} 
\edef\MIT{$^{\arabic{univ_counter}}$ } 

\addtocounter{univ_counter} {1} 
\edef\UMASS{$^{\arabic{univ_counter}}$ } 

\addtocounter{univ_counter} {1} 
\edef\MSU{$^{\arabic{univ_counter}}$ } 

\addtocounter{univ_counter} {1} 
\edef\UNH{$^{\arabic{univ_counter}}$ } 

\addtocounter{univ_counter} {1} 
\edef\NSU{$^{\arabic{univ_counter}}$ } 

\addtocounter{univ_counter} {1} 
\edef\ODU{$^{\arabic{univ_counter}}$ } 

\addtocounter{univ_counter} {1} 
\edef\PITT{$^{\arabic{univ_counter}}$ } 

\addtocounter{univ_counter} {1} 
\edef\RPI{$^{\arabic{univ_counter}}$ } 

\addtocounter{univ_counter} {1} 
\edef\RICE{$^{\arabic{univ_counter}}$ } 

\addtocounter{univ_counter} {1} 
\edef\URICH{$^{\arabic{univ_counter}}$ } 

\addtocounter{univ_counter} {1} 
\edef\UCS{$^{\arabic{univ_counter}}$ } 

\addtocounter{univ_counter} {1} 
\edef\VIRGINIA{$^{\arabic{univ_counter}}$ } 

\addtocounter{univ_counter} {1} 
\edef\VT{$^{\arabic{univ_counter}}$ } 

\addtocounter{univ_counter} {1} 
\edef\WM{$^{\arabic{univ_counter}}$ } 

\addtocounter{univ_counter} {1} 
\edef\YEREVAN{$^{\arabic{univ_counter}}$ } 

%%%%%%%%%%%%%%%%%%%%%%%%%%%%%%%%%  TITLE %%%%%%%%%%%%%%%%%%%%%%%%%%%%%%%%
\title{ Observation of an Exotic $S=+1$ Baryon in 
Exclusive Photoproduction from the Deuteron}

%%%%%%%%%%%%%%%%%%%%%%%%%%%%%%%  AUTHORS %%%%%%%%%%%%%%%%%%%%%%%%%%%%%%%% 

\author{ 
S.~Stepanyan,\JLAB$\!\!^,$\ODU\
K.~Hicks,\OHIOU\
D.S.~Carman,\OHIOU\
E.~Pasyuk,\ASU\
R.A.~Schumacher,\CMU\
E.S.~Smith,\JLAB\
D.J.~Tedeschi,\SCAROLINA\
L.~Todor,\CMU\
G.~Adams,\RPI\
P.~Ambrozewicz,\FIU
% ...
\\
(CLAS collaboration)
} 

\affiliation{\JLAB Thomas Jefferson National Accelerator Laboratory, Newport News, Virginia 23606}
\affiliation{\OHIOU Ohio University, Athens, Ohio  45701}
\affiliation{\ASU Arizona State University, Tempe, Arizona 85287}
\affiliation{\CMU Carnegie Mellon University, Pittsburgh, Pennsylvania 15213}
\affiliation{\SCAROLINA University of South Carolina, Columbia, South Carolina 29208}
\affiliation{\SACLAY CEA-Saclay, DAPNIA-SPhN, F91191 Gif-sur-Yvette Cedex, France}
\affiliation{\UCLA University of California at Los Angeles, Los Angeles, California  90095}
\affiliation{\CUA Catholic University of America, Washington, D.C. 20064}
\affiliation{\CNU Christopher Newport University, Newport News, Virginia 23606}
\affiliation{\UCONN University of Connecticut, Storrs, Connecticut 06269}
\affiliation{\DUKE Duke University, Durham, North Carolina 27708}
\affiliation{\EDINBURGH Edinburgh University, Edinburgh EH9 3JZ, United Kingdom}
\affiliation{\FIU Florida International University, Miami, Florida 33199}
\affiliation{\FSU Florida State University, Tallahassee, Florida 32306}
\affiliation{\GWU The George Washington University, Washington, DC 20052}
\affiliation{\GLASGOW University of Glasgow, Glasgow G12 8QQ, United Kingdom}
\affiliation{\INFNFR INFN, Laboratori Nazionali di Frascati, P.O. 13,00044 Frascati, Italy}
\affiliation{\INFNGE INFN, Sezione di Genova and Dipartimento di Fisica, Universit\`a di Genova, 16146 Genova, Italy}
\affiliation{\ORSAY Institut de Physique Nucleaire d'ORSAY, IN2P3, BP1, 91406 Orsay, France}
\affiliation{\ITEP Institute of Theoretical and Experimental Physics, Moscow, 117259, Russia}
\affiliation{\JMU James Madison University, Harrisonburg, Virginia 22807}
\affiliation{\KYUNGPOOK Kyungpook National University, Taegu 702-701, South Korea}
\affiliation{\MIT Massachusetts Institute of Technology, Cambridge, Massachusetts  02139}
\affiliation{\UMASS University of Massachusetts, Amherst, Massachusetts  01003}
\affiliation{\MSU Moscow State University, 119899 Moscow, Russia}
\affiliation{\UNH University of New Hampshire, Durham, New Hampshire 03824}
\affiliation{\NSU Norfolk State University, Norfolk, Virginia 23504}
\affiliation{\ODU Old Dominion University, Norfolk, Virginia 23529}
\affiliation{\PITT University of Pittsburgh, Pittsburgh, Pennsylvania 15260}
\affiliation{\RPI Rensselaer Polytechnic Institute, Troy, New York 12180}
\affiliation{\RICE Rice University, Houston, Texas 77005}
\affiliation{\URICH University of Richmond, Richmond, Virginia 23173}
\affiliation{\UCS Union College, Schenectady, New York 12308}
\affiliation{\VIRGINIA University of Virginia, Charlottesville, Virginia 22901}
\affiliation{\VT Virginia Polytechnic Institute and State University, Blacksburg, Virginia 24061}
\affiliation{\WM College of William and Mary, Williamsburg, Virginia 23187}
\affiliation{\YEREVAN Yerevan Physics Institute, 375036 Yerevan, Armenia}

\date{\today}

%%%%%%%%%%%%%%%%%%%%%%%%%%%%%%% ABSTRACT %%%%%%%%%%%%%%%%%%%%%%%%%%%%%%%%%

\vfil

\begin{abstract}

In an exclusive measurement of the reaction $\gamma d \rightarrow K^+
K^- p n$, a narrow peak that can be attributed to an exotic baryon
with strangeness $S=+1$ is seen in the $K^+n$ invariant mass
spectrum. The peak is at $1.542\pm 0.005$ GeV/c$^2$ with a measured width
of 0.021 GeV/c$^2$ FWHM, which is largely determined by experimental mass
resolution. The statistical significance of the peak is $5.2 \pm 0.6\
\sigma$. The mass
and width of the observed peak are consistent with recent reports of a narrow
$S=+1$ baryon by other experimental groups.

\end{abstract}


\maketitle
\newpage

High-energy neutrino and anti-neutrino 
scattering experiments \cite{perkins} have established that sea-quarks
($q\bar{q}$ pairs)  
are part of the ground-state wave function of the nucleon. 
In addition, results from pion electroproduction experiments in the $\Delta$-resonance 
region, together with other experiments, have shown \cite{csmith} the
presence of a pion ``cloud" surrounding the valence quarks of the
nucleon. In this sense, 
5-quark ($qqqq\bar{q}$) configurations are mixed with the standard 
3-quark valence configuration.  However, it is natural to ask whether a 5-quark 
configuration exists where the $\bar{q}$ has a different flavor than 
(and hence cannot annihilate with) the other four quarks.  
Such states are not forbidden by QCD \cite{strottman,lipkin}, 
and definite evidence of pentaquark states would be an important 
addition to our understanding of QCD. In fact, the question of 
which color singlet configurations exist in nature lies at the 
heart of non-perturbative QCD.  A baryon with the 
exotic strangeness quantum number $S=+1$ is a natural candidate 
for a pentaquark state.

The general idea of a five-quark state has been around since the late
60's \cite{pdg86}. Recently, 
symmetries within the chiral soliton model were used by 
Diakonov, Petrov and Polyakov \cite{diakonov} 
to predict an anti-decuplet 
of 5-quark resonances with spin and parity $J^\pi = \frac{1}{2}^+$. 
The lowest mass member, an isosinglet with valence quark 
configuration $uudd\bar{s}$ giving strangeness $S=+1$ 
(originally called the $Z^+$, but now renamed 
as the $\Theta^+$\cite{thname}) 
has a predicted mass of approximately 1.53 GeV/c$^2$ and a 
width of $\sim 0.015$ GeV/c$^2$. 
The narrow width, similar to that of the $\Lambda$(1520) 
baryon resonance with strangeness $S=-1$, is largely constrained 
by symmetries of the coupling constants and 
the phase space of the decay to a $KN$ final state.

The existence of the \thp has been suggested by several recent experiments.  
The LEPS collaboration at the SPring-8 facility in Japan recently 
reported \cite{nakano} the observation of an $S=+1$ baryon at 
1.54 GeV/c$^2$ with a FWHM less than $0.025$ GeV/c$^2$ from the inclusive 
reaction $\gamma n \to K^-(K^+X)$ where the target neutron is bound in 
carbon,  and
the residual nucleus is assumed to be a spectator.
This measurement reported a statistical 
significance of $4.6 \pm 1.0\ \sigma$.  Also,
the DIANA collaboration at ITEP \cite{dolgolenko} recently 
announced results from an analysis of bubble-chamber data 
for the reaction $K^+ n \rightarrow K^0 p$, where the neutron 
is bound in a xenon nucleus, which shows a narrow peak at 
$1.539 \pm 0.002$ GeV/c$^2$.
The statistical significance of the ITEP result is 4.4 $\sigma$. 

One might wonder how an $(nK^+)$ resonance could have evaded earlier 
searches.  The Particle Data Group has summarized these searches 
most recently in 1986 \cite{pdg86} saying that the results permit 
no definite conclusion for an $S=+1$ resonance. A more recent 
phase shift analysis for $K^+N$ scattering by Hyslop {\it et al.} 
\cite{hyslop} finds weak resonance behavior in 
the $P_{01}$, $D_{03}$, $P_{13}$, and $D_{15}$ partial waves, 
but none of these ``resonance-like structures" (in the words of 
Ref. \cite{hyslop}) have convincing phase motion.  We note 
that the $K^+n$ database is sparse for low-energy $K^+$ beams, 
and a narrow resonance could have escaped detection \cite{arndt}.

The present experiment is an exclusive measurement on deuterium 
for the reaction $\gamma d \rightarrow K^+ K^- p (n)$ where the final
state neutron is reconstructed from the missing momentum and energy.  
The data presented here were taken at the Thomas Jefferson National 
Accelerator Facility with the CLAS detector \cite{clas}
and the photon tagging system \cite{tagger} in Hall B.
Photon beams were produced by
2.474 and 3.115 GeV electrons incident on a bremsstrahlung 
radiator of thickness $10^{-4}$ radiation lengths, giving a 
tagged photon flux of approximately $4\times 10^6$ $\gamma$'s per
second. The maximum tagged photon energy was $95\%$ of the electron
beam energy.
The integrated tagged photon flux above $1.51$ GeV was $1.64\times 10^{12}$ at 2.474 GeV 
and $0.70\times 10^{12}$ at 3.115 GeV.
The tagged photon energy is measured with a resolution of between 
0.003 and 0.005 GeV, depending on the energy.
The photons struck a liquid-deuterium target of thickness 10 cm. 

%%%%%%%%%%%%%%%%%%%%%% Figure : Neutron Mass %%%%%%%%%%%%%%%%%%%%%%%%%%%%%
\begin{figure}[htb]
\includegraphics[width=1.\columnwidth]{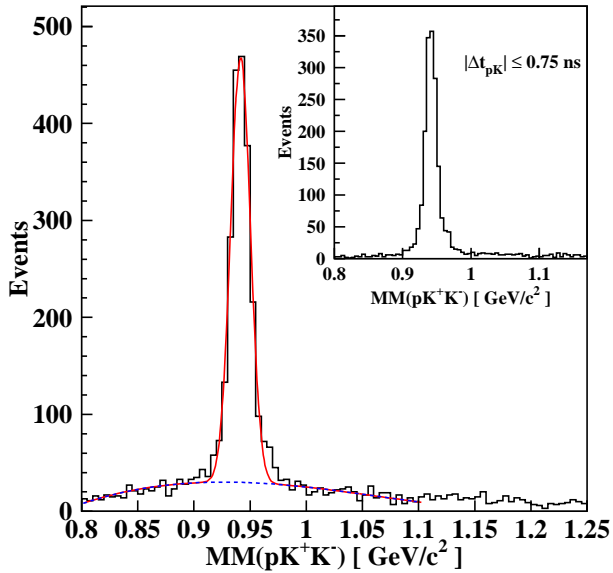}
\caption{ Missing mass spectrum for the $\gamma d \rightarrow pK^+K^- X$ 
reaction, after timing cuts to identify the charged particles and 
the coincident photon, which shows a peak at the neutron mass. 
There is a small, broad background from misidentified particles 
and other sources.  The inset shows the neutron peak with a 
tighter requirement on the timing between the proton and kaons.
}
\label{fig:neutron}
\end{figure}
%%%%%%%%%%%%%%%%%%%%%%%%%%%%%%%%%%%%%%%%%%%%%%%%%%%%%%%%%%%%%%%%%%%%%%%%%%%%%%

The event trigger was formed when a charged particle hit two
scintillator planes (the ``start counter" around the target 
and a time-of-flight ``TOF" counter a few meters away), 
in coincidence with an electron detected in the tagging system.  
The particle identification was performed using the reconstructed 
momentum and charge from tracking, 
together with the measured TOF.
The analysis focused on events with a detected proton, 
$K^+$ and  $K^-$ (and no other charged particles) in the final state.
Either the $K^+$ or the $K^-$ in the event was required to have a
time at the interaction vertex within 1.5 ns of the proton's vertex time. 
Also, the incident photon time at the interaction vertex was required to be within 
1.0 ns of the proton (to eliminate accidental coincidences).
The missing mass ($MM$) of the selected events is plotted 
in Fig.~\ref{fig:neutron}, which shows a peak at 
the neutron mass on top of a small background. A fit to the
distribution (solid line) yields 
a mass resolution of $\sigma = 0.009$ GeV/c$^2$. 

%%%%%%%%%%%%%%%%%%%%%% Figure : Feynman Diagrams %%%%%%%%%%%%%%%%%%%%%%%%%%%%%
\begin{figure}[htb]
\includegraphics[width=0.8\columnwidth]{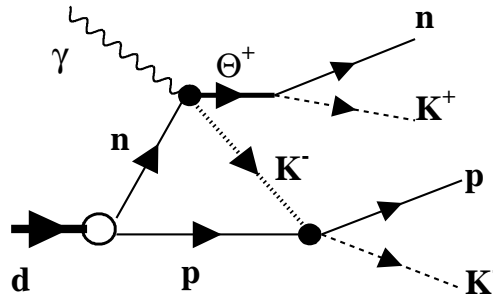}
\caption{ A rescattering diagram that could contribute to the exclusive 
reaction mechanism that produces the \thp and an energetic proton 
through final state interactions. Note that the \thp is produced 
independently of the secondary scattering.
}
\label{fig:diagrams}
\end{figure}
%%%%%%%%%%%%%%%%%%%%%%%%%%%%%%%%%%%%%%%%%%%%%%%%%%%%%%%%%%%%%%%%%%%%%%%%%%%%%%

The reaction $\gamma d \rightarrow K^+ K^- p (n)$ selects the \thp
decays to the $K^+n$ final state. 
It is likely that production of the \thp in this final state proceeds
on the neutron via $t$-channel $K^+$ exchange, similar to the
production of the $\Lambda(1520)$ on the proton, where
the dominant mechanism is $t$-channel $K^-$ exchange \cite{lambda}.   
If the proton is a spectator during 
\thp production, it will not be seen in our detector due to 
its small momentum \cite{minmom}.  
However, in some fraction of events, the
$K^-$ and the proton may be involved in the final state interaction
(FSI), as shown in Fig.\,\ref{fig:diagrams}. While the production of 
the \thp does not require a rescattering,
such events increase the probability of detecting the $K^-$'s and the
protons in the final state by rescattering them into the acceptance
region of CLAS. By requiring an exclusive process, we are able
to fully reconstruct the unobserved neutron, which 
aids significantly in reducing background.
Another advantage of this exclusive measurement is 
that no corrections for the Fermi momentum of 
the neutron are needed, so the width of the neutron peak 
in Fig. \ref{fig:neutron} is limited only 
by the resolution of the CLAS detector system.  Events 
within $\pm 3\ \sigma$ of the neutron peak were 
kept for further analysis.  The background in this region 
is about 15\% of the total, mostly from pions 
that are misidentified as kaons.

%%%%%%%%%%%%%%%%%%%%%% Figure : Phi and Lambda* %%%%%%%%%%%%%%%%%%%%%%%%%%%%%
\begin{figure}[htb]
\includegraphics[width=1.\columnwidth]{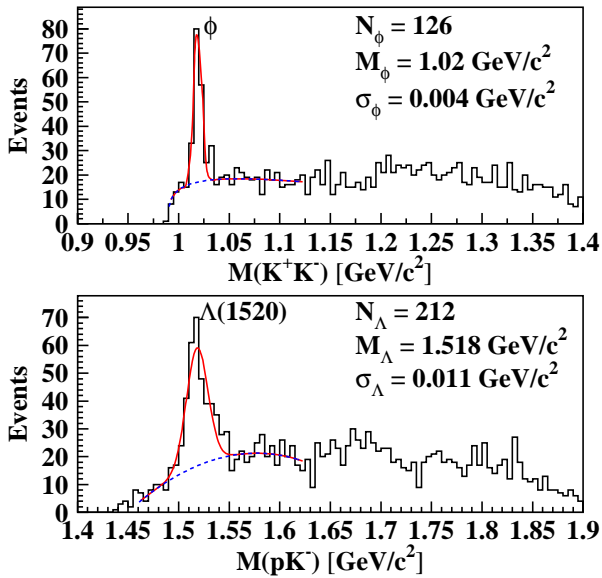}
\caption{ Invariant mass of the $K^+K^-$ system (top) and the 
$pK^-$ system (bottom) showing peaks at the mass of known resonances. 
These resonances are removed in the analysis by placing cuts on 
the peaks shown.  Results for the number of counts (N), the mass (M) 
and the widths ($\sigma$) from fits are also given.
}
\label{fig:philam}
\end{figure}
%%%%%%%%%%%%%%%%%%%%%%%%%%%%%%%%%%%%%%%%%%%%%%%%%%%%%%%%%%%%%%%%%%%%%%%%%%%%%%

There are several known reactions, such as photoproduction of
mesons (that decay into $K\bar{K}$) or excited hyperons (that 
decay into $pK^-$), that contribute to the same final state.  
We now discuss the explicit cuts we have made to remove the main
background sources from our final event sample in order to enhance
our signal relative to background.
The $\phi$ meson at 1.02 GeV/c$^2$, which decays into a $K^+$ and $K^-$,
is produced primarily at forward angles \cite{betsch}. 
These events are easily identified using the invariant mass 
of the $K^+K^-$ pair, $M(K^+K^-)$, as shown in 
Fig. \ref{fig:philam} (top).
In order to remove the $\phi$ mesons, events with 
$M(K^+K^-) < 1.07$ GeV/c$^2$ are rejected.

Similarly, the $\Lambda$(1520) resonance can be produced
by the $\gamma p \rightarrow K^+ \Lambda$ reaction
with subsequent decay of the $\Lambda$ to a proton and $K^-$.  
A peak corresponding to the $\Lambda$(1520) is seen in the invariant 
mass spectrum of the $pK^-$ system, $M(pK^-)$, as shown in 
Fig. \ref{fig:philam} (bottom). Unlike $\phi$ mesons,
$\Lambda$(1520)'s can be produced in conjunction with $\Theta^+$'s and still
conserve net strangeness. However, even though there is a large
cross-section for producing $\Lambda$(1520)$K^+$ on the proton
followed by $K^+n$
rescattering, the kinematics is a
poor match for \thp production, since, as was described above, this
is likely a $t$-channel process with forward production of the $K^+$ in the
CM frame. In our kinematics the average momentum for the $K^+$ in the
production of the $\Lambda$(1520) is $\sim 0.8$
GeV/c, while for the production of the $\Theta^+$ in $K^+n$
interaction, the average momentum of the kaon should be approximately
$0.45$ GeV/c. For this reason, we reject
events with 1.485$<M(pK^-)<$1.551 GeV/c$^2$ ($\pm
3\sigma$ cut from the peak) to improve our signal to
background ratio.
% even at the expense of a small loss of signal. 

Two other event selection requirements were applied, based 
on kinematics.  
The first one requires that the missing momentum of the 
undetected neutron must be greater than 0.08 GeV/c.  Below this 
value, the neutron is likely a spectator to other reaction 
mechanisms.  Our studies show that increasing the 
value of this cutoff does not change the final results -- in 
particular it does not eliminate the peak shown below -- but
does reduce the statistics in the $M(nK^+)$ spectrum.
The second requirement concerns the $K^+$ momentum.  
Monte Carlo simulations of the 
\thp decay from an event distribution uniform in phase space 
show that the $K^+$ momentum rarely exceeds 1.0 GeV/c.  The 
data also show that $K^+$ momenta greater than 1.0 GeV/c are 
associated with an invariant mass of the $nK^+$ system 
above $\sim 1.7$ GeV/c$^2$.  Events with a $K^+$ 
momentum above 1.0 GeV/c were removed to reduce this background.

%%%%%%%%%%%%%%%%%%%%%% Figure : Final Theta+ %%%%%%%%%%%%%%%%%%%%%%%%%%%%%
\begin{figure}[htb]
\includegraphics[width=1.\columnwidth]{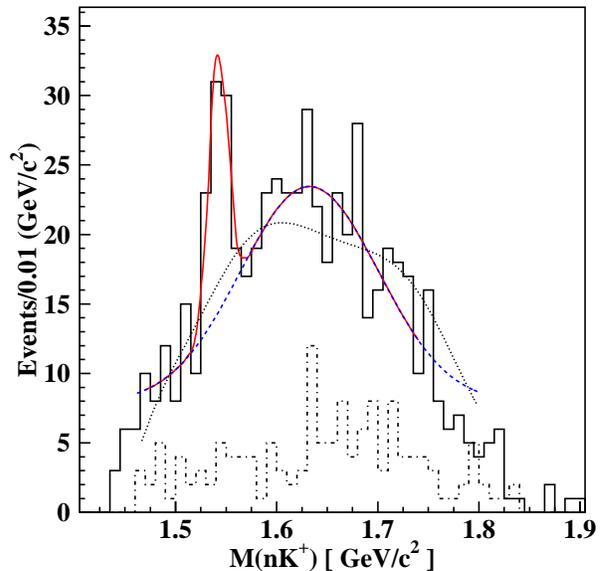}
\caption{ Invariant mass of the $nK^+$ system, which has strangeness 
$S=+1$, showing a sharp peak at the mass of 1.542 GeV/c$^2$.  
A fit (solid line) to the peak on top of the smooth background
(dashed line)
gives a statistical significance of 5.8 $\sigma$. 
The dotted curve is the shape of the simulated background. 
The dashed-dotted histogram shows the spectrum of events 
associated with $\Lambda$(1520) production. 
}
\label{fig:final}
\end{figure}
%%%%%%%%%%%%%%%%%%%%%%%%%%%%%%%%%%%%%%%%%%%%%%%%%%%%%%%%%%%%%%%%%%%%%%%%%%%%%%

The final $nK^+$ invariant mass spectrum, $M(nK^+)$, 
is shown in Fig.\,\ref{fig:final} \cite{acccor}, along with a fit
(solid line) to the peak and a Gaussian plus constant term fit 
to the background (dashed line). 
For the fit given, there are 43 counts in the peak at a 
mass of $1.542 \pm 0.005$ GeV/c$^2$ with a width (FWHM) of 
0.021 GeV/c$^2$.  The width is consistent with  
the instrumental resolution.  The uncertainty of 0.005 GeV/c$^2$ in 
the mass is due to calibration uncertainties of the 
photon tagging spectrometer  
\cite{tagger}, the electron beam energy, and the 
momentum reconstruction in CLAS.
The statistical significance of 
this peak is estimated based on fluctuations 
of the background over a $\pm 2\sigma$ window centered 
on the peak, giving $43 / \sqrt{54}=5.8\ \sigma$. 
The spectrum of events removed by the
$\Lambda$(1520) cut is shown in Fig.\,\ref{fig:final} by the dashed-dotted
histogram, and does not appear to 
be associated with the peak at 1.542 GeV/c$^2$.

The shape of the expected $M(nK^+)$ mass spectrum was investigated by 
a Monte Carlo simulation using GEANT \cite{geant} based simulation
tools for the CLAS detector and the algorithm used for the data
analysis. We studied 4-body phase space production of the $pK^+K^-n$
final state, and the production of the  
3-body phase space in the $pK^+K^-$ final
state ($K^+K^-$ in s-wave). 
No peak-like structures were visible in the $M(nK^+)$
distributions of these two final states.
We used the shapes of these distributions
to fit the experimental $M(nK^+)$ spectrum.
The fitted shape of the background is shown by the dotted line in
Fig. \ref{fig:final}. The relative weights of 3-body and 4-body
phase space events determined by the fit was $3:1$. The statistical significance
of the peak at $1.542$ GeV/c$^2$ in the fit using this simulated background
was $4.8 \sigma$. 

A separate Monte Carlo study was carried out to examine 
the production of known resonances via the reaction 
$\gamma d \rightarrow K^+Y^*N$, where the $Y^*$ decays 
to a $K^-N$ followed by one of the kaons rescattering 
off the spectator nucleon.  This study \cite{meyer} was 
unable to produce structures narrower than about
four times the CLAS resolution, and concluded that these 
rescattering processes are not responsible for such a 
narrow structure in the $M(nK^+)$ spectrum.

\begin{figure}[htb]
\includegraphics[width=1.\columnwidth]{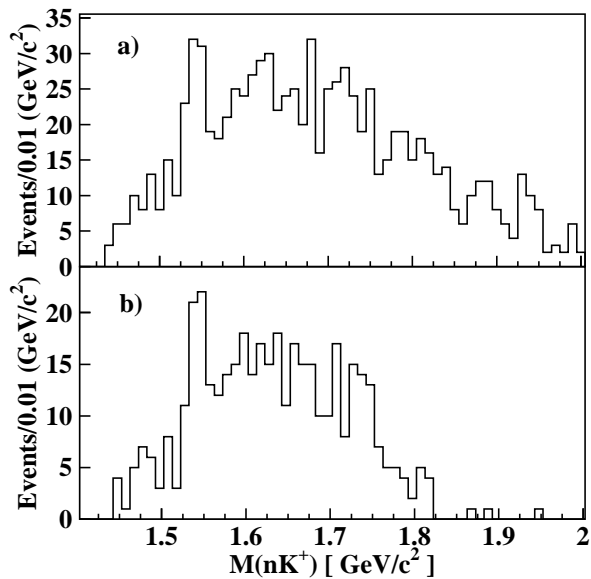}
\caption{ Spectra of the $(nK^+)$ invariant mass for different cuts:
%with the same cuts as
%in Fig.\,\ref{fig:final} with the following exceptions:
a) no cut on the $K^+$ momentum;
b) tight cut on proton-kaon vertex time. See text for details.}
%each kaon is required 
%to be within 0.75 ns of the proton's time at the interaction vertex.
%}
\label{fig:varycuts}
\end{figure}

The sensitivity of the peak to the placement of event selection 
cuts was studied, and the conclusion is that the peak at 
1.542 GeV/c$^2$ is very robust.  For example, removing the 
$K^+$ momentum limit 
results in the spectrum shown in Fig.\,\ref{fig:varycuts}a. 
Alternatively, 
tightening the cuts on proton-kaon timing, $|\Delta t_{pK}|<0.75$ ns,
allows less background  
into the spectrum, as shown in the inset of Fig. \ref{fig:neutron}.
The shape of the $M(nK^+)$ spectrum for this selection is shown in
Fig.~\ref{fig:varycuts}b and  
remains essentially unchanged from Fig.\,\ref{fig:final}. 
In all we tried 10 variations of event
cut placement and/or different fitting functions. All fits reproduce
the measured data with reduced $\chi^2$ in the range between 0.6 and
1. The estimated statistical significance in those ten cases ranges from
$4.6 \sigma$ to $5.8 \sigma$, which we use to derive the conservative
estimate for the statistical significance of our result of $5.2 \pm
0.6\ \sigma$. 

%%%%%%%%%%%%%%%%%%%%%% Figure : other M(nK+) %%%%%%%%%%%%%%%%%%%%%%%%%%%%%
%%%%%%%%%%%%%%%%%%%%%%%%%%%%%%%%%%%%%%%%%%%%%%%%%%%%%%%%%%%%%%%%%%%%%%%%%%%%%%

A neutron and $K^+$ can couple to both isospin zero and isospin one
states.	If the $\Theta^+$ has $I=1$, then there should be two other
members of the isotriplet, a neutral and a doubly charged state. The
doubly charged state would couple to $pK^+$.
We examined 
the invariant mass $M(pK^+)$ using the same event selection 
as before.  The statistics are limited, but there is no clear 
peak in the signal region.  It should be noted that the CLAS 
acceptance for the $pK^+$ system is not the same as for 
the $nK^+$ system, so the two spectra are not directly comparable.  
The featureless $M(pK^+)$ spectrum (not shown) suggests that the peak 
at 1.542 GeV/c$^2$ in the $M(nK^+)$ spectrum is an isosinglet, 
but it is difficult to draw a firm conclusion based on the 
current data.

These results from CLAS, together with other experiments  
\cite{nakano,dolgolenko}, now 
provide convincing evidence for the existence of an $S$=+1 
baryon state at a mass of $1.542$ GeV/c$^2$ with a small intrinsic 
width. In this paper we presented evidence for this state with a
statistical significance in the  
range of $5.2 \pm 0.6\ \sigma$, depending on estimates of the 
background and on the event selection criteria. However, more studies
are needed before  
this $S=+1$ state can be conclusively identified 
with the \thp predicted in Ref.~\cite{diakonov}.  Further evidence for
the \thp should be searched for in a variety 
of reactions, in addition to the ones mentioned here.  
Spin, isospin, and parity of this state remain to be 
established in future experiments.

We would like to acknowledge the outstanding efforts of the 
staff of the Accelerator Division and the Physics Division at Jefferson 
Lab that made this experiment possible.  This work is supported 
by the U.S. Department of Energy and the National Science Foundation, 
the French Commissariat \`a l'Energie Atomique, the French Centre
National de la Recherche Scientifique, the Italian Istituto 
Nazionale di Fisica Nucleare, the Korea Science and Engineering 
Foundation, and the UK Engineering and Physical Research Sciences Council. The
Southeastern Universities Research Association 
(SURA) operates the Thomas Jefferson National Accelerator Facility 
for the U.S. Department of Energy under contract DE-AC05-84ER40150.


\begin{thebibliography}{99}
\bibitem{minmom} The minimum momentum of charged particles detected
in CLAS is angle dependent \cite{clas}, and approximately $0.3$ GeV/c
for the kinematics of this reaction..
\bibitem{acccor} None of presented mass
distributions are acceptance corrected.
\end{thebibliography}
\end{document}